Submitted to 'Chinese Physics C'

# The Development of ADS Virtual Accelerator Based on XAL

Wang Peng-Fei (王鹏飞)[1,2] Cao Jian-She (曹建社)[1] Ye Qiang (叶强)[1]

[1] Institute of High Energy Physics, Chinese Academy of Sciences, Beijing 100049, China

[2] University of Chinese Academy of Sciences, Beijing 100049, China

*Abstract*

XAL [1-3] is a high level accelerator application framework originally developed by the Spallation Neutron Source (SNS), Oak Ridge National Laboratory. It has advanced design concept and adopted by many international accelerator laboratories. Adopting XAL for ADS is a key subject in the long term. This paper will present the modifications to the original XAL applications for ADS. The work includes proper relational database schema modification in order to better suit ADS configuration data requirement, redesigning and re-implementing db2xal application and modifying the virtual accelerator application. In addition, the new device types and new device attributes for ADS online modeling purpose is also described here.

## 1. Background

The Accelerator Driven Sub-critical System (ADS) takes the spallation neutrons as the external neutron source to drive the sub-critical blanket system [4]. So, it has the inherent safety and has been universally regarded as the most effective approach to dispose the long-lived nuclear wasted [5]. In 2011, the Chinese Academy of Sciences launched the "Strategic Priority Research Program" named "Future Advanced Nuclear Fission Energy" [4]. This program has two sub-programs, and the ADS Project is one of them.

XAL is a mature framework for rapid development of applications. It is written in Java, and provides users with a hierarchal view of the accelerator which shown schematically in Fig.1. Features include database configuration of the accelerator structure, a common look-and-feel graphical user interface (GUI), an online envelope model that is configurable from design or live machine values, a scripting interface for algorithm development, and many other utility packages.

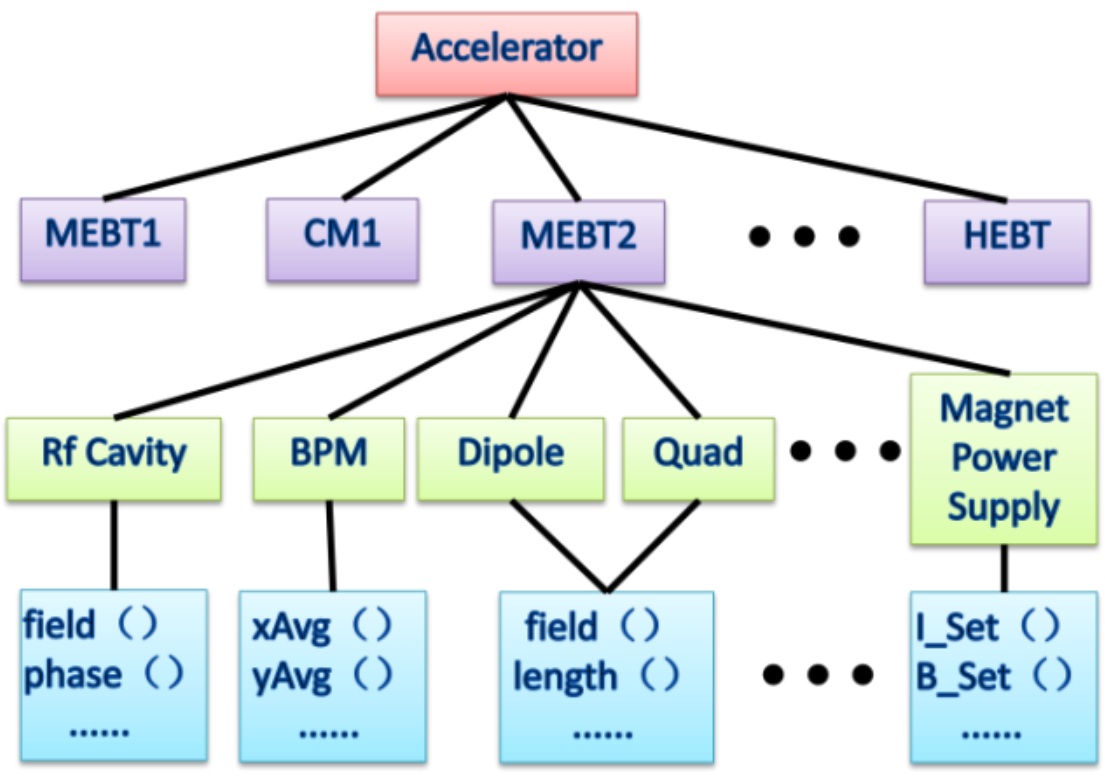

Fig.1: Schematic of the XAL accelerator class hierarchy.

## 2. ADS accelerator database

The decision to use MySQL was made easily and early: the world's most popular open source database; it is widely used in the world by many of the largest organizations including Facebook, Google and Adobe; it is extremely easy to use; scalability and flexibility; high performance; high availability; robust transactional support; strong data protection; management ease; the relational model

wangpf@ihep.ac.cn

is well known and understood. Compared with relational database, OO databases have some stability-related "early adopter" problems. These factors led us to choose MySQL.

The ADS Virtual Accelerator Database that shown in Fig.2 use the standard relational model. It is redesigned and implemented by referencing SNS [6] and CSNS [7] Virtual Accelerator Database

## 2.1 Database brief introduction

1) BEAM_LINE_DVC: Information of beam line devices such as the measured misalignments, aperture shape and size.
2) DVC: Information of all devices including standby equipment.
3) BEAM_LINE: Version information of beam line.
4) DVC_WS_SFTW: The WS device is profile monitor or wire scanner.
5) DVC_SET: Information of the device settings for beam line devices.
6) CHANNEL: The master list of valid EPICS PVs and its handler that classified by device type.
7) BPM_DVC: Unique information of the BPM device.
8) DVC_SEQ: Information of accelerator sequences.
9) RF_GAP: Unique information of the RF cavity gap.
10) RF_DVC: Unique information of the RF cavity.
11) MAG_DVC: Unique information of the magnet.

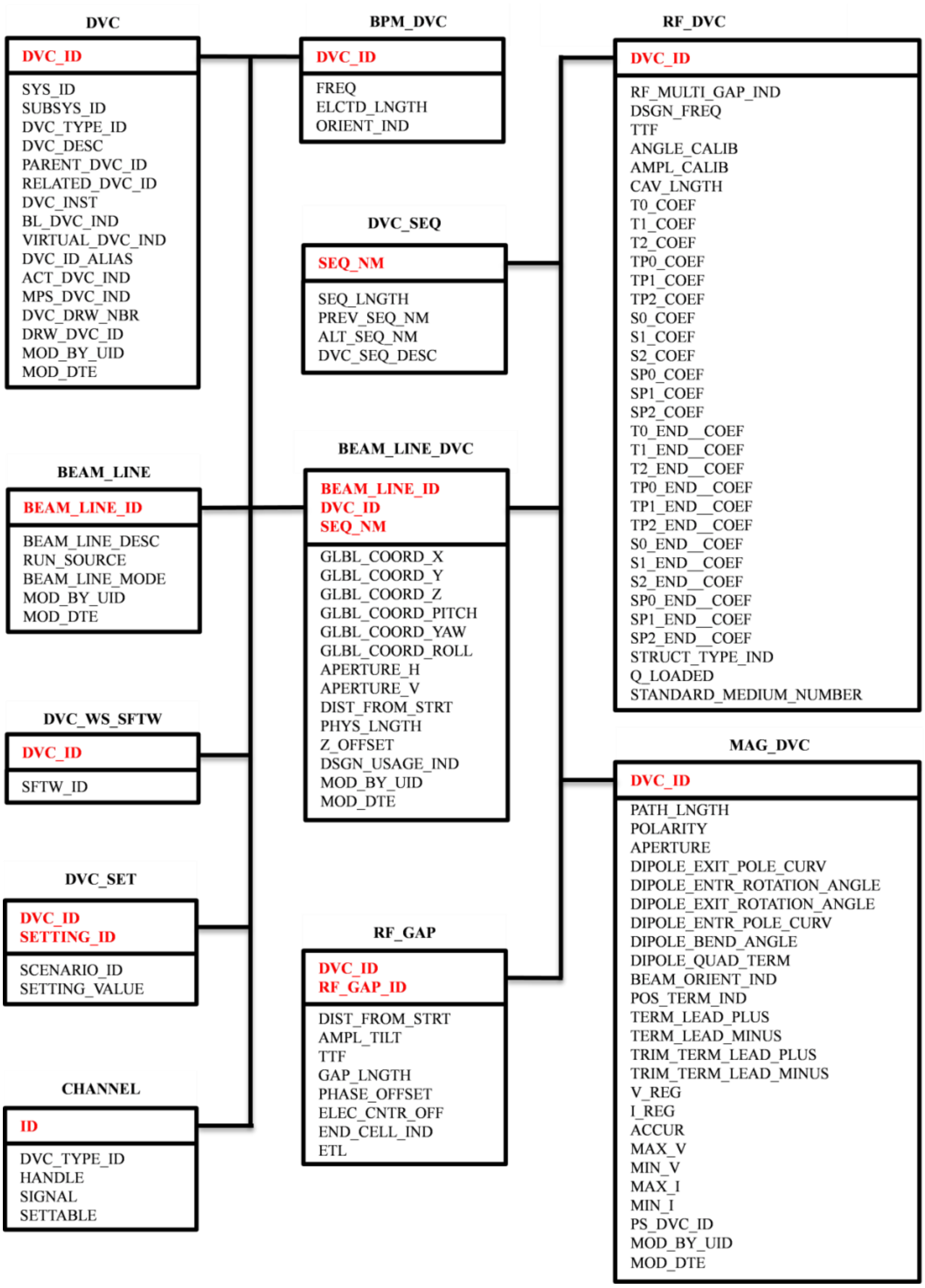

Fig.2: ADS Virtual Accelerator Database

## 3. ADS virtual accelerator configuration files

The XAL configuration mechanism is driven by five main files and two auxiliary files, as shown in Fig.3. The main.xal file contains the names and locations of other four main files, which are: ads.impl, ads.xdxf, model.params, timing_pvs.tim. The two auxiliary files are ADSMEBT1Entr.probe and xdxf.dtd.

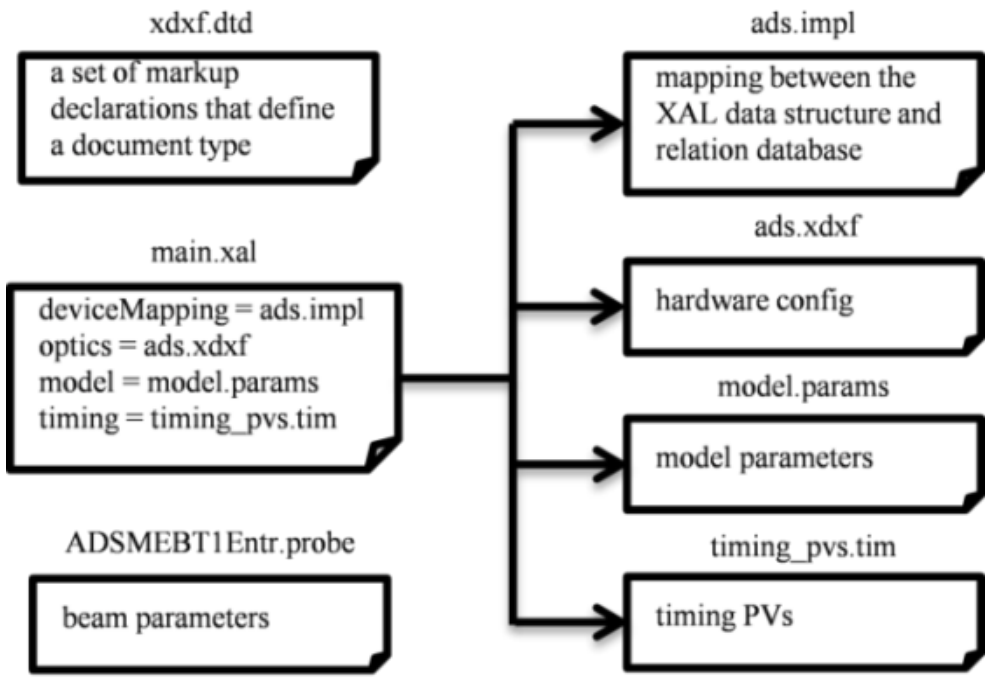

Fig.3: XAL configuration files

## 4. Db2xal

Among the XML configuration files, the ads.xdxf file has tens of thousands of lines. It is difficult to manually generate this file. Db2xal is a XAL general purpose application to do this tedious task, but the original db2xal has much SNS specific information embedded in the code which leads to bad performance in ADS. ADS also has specific elements, like the solenoid element. Besides, the bunchers in MEBT1 and RF cavities in CM1 use the new attribute ETL instead of the separate properties—the longitudinal electric field(E), the transit time factor (T) and the gap length (L). In addition, the original db2xal application has poor interaction**.** Redesigning and re-implementing db2xal application to better suit ADS configuration data requirement is needed. The flowchart of db2xal is shown in Fig.4.

### 4.1 Implementing

1) Three main classes

• Main. It is a subclass of ApplicationAdaptor base class.

• Db2XalDocument. As a general purpose application, the Db2XalDocument class subclass the XalDocument base class. It typically implements the actions for the views defined in the gui.bricks file and the menu definitions file.

• Db2XalExtractDataFromDB. It is the main class responsible for extracting data from the database to generate corresponding optics configuration information.

2) Four auxiliary classes

• Db2XalMyTableModel. It is an inner class of Db2XalDocument and used to generate data of the table shown in the left part of Fig.5 and handle the button event

• MyTimerTask. It is an inner class of Db2XalDocument and used to update the information of the status bar and the status of the progress bar

• Db2XalBeamlineIDTableDialog. It is used to pop up a dialog box and handle the button event

• Db2XalBeamlineIDTableModel. It is an inner class of Db2XalBeamlineIDTableDialog and used to generate the data of table in dialog box

3) GUI file [8]

• Gui.bricks which resides in the “resources” subfolder of Db2XAL is generated by the Bricks application. It describes the views within the document’s associated main window. The main interface of db2xal include title bar, menu bar, toolbar, workspace and status bar shown in Fig.5. The title bar and toolbar is provided by the XAL framework. The menu bar is defined in the menu definition file which resides in the same folder of Gui.bricks. Gui.bricks describes the layout of the views within the workspace and status bar.

### 4.2 Program function

The ADS initial design development stage wherein, it is possible that errors will be discovered which will require design changes. The version information of beam line such as beam line ID and beam line info is used to record these changes. The read-only textbox shown in the upper part of Fig.5 is used to display the version information. Beam line ID can be changed by click the “Change ID” button. In order to generate configuration file, it is needed to select appropriate beam line ID and continuous sequences. A warning message—“You must select continuous sequences!” appears when discontinuous sequences are selected. When the user click the “Generate XDXF” button, the main process of db2xal will extract data from development database or production database and display the result in the textarea control. Development database that mainly used by program staffs is used to test new device or sequence. Production database as the db2xal’s default database is used to daily operations. User can change the default database by selecting the “Database” menu item. Meanwhile, some measures are taken to promote the interactivity such as using a status label to display messages during different phases of the application’s life cycle, using a progress bar to show an approximate percentage of completion of the main process. The final result displayed in textarea control can be edited according to customer needs or be exported to “ads.xdxf” should the user click the “Save” button.

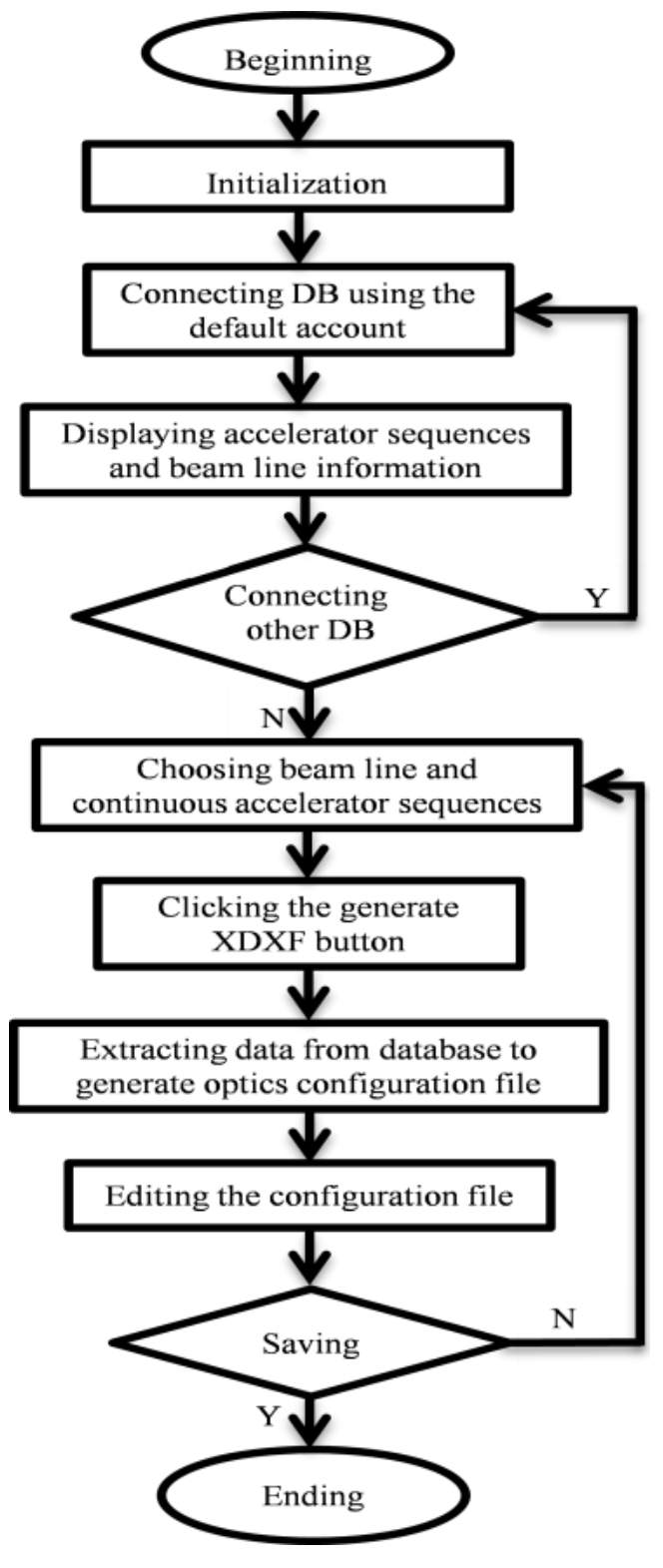

Fig.4: Flowchart of db2xal [10]

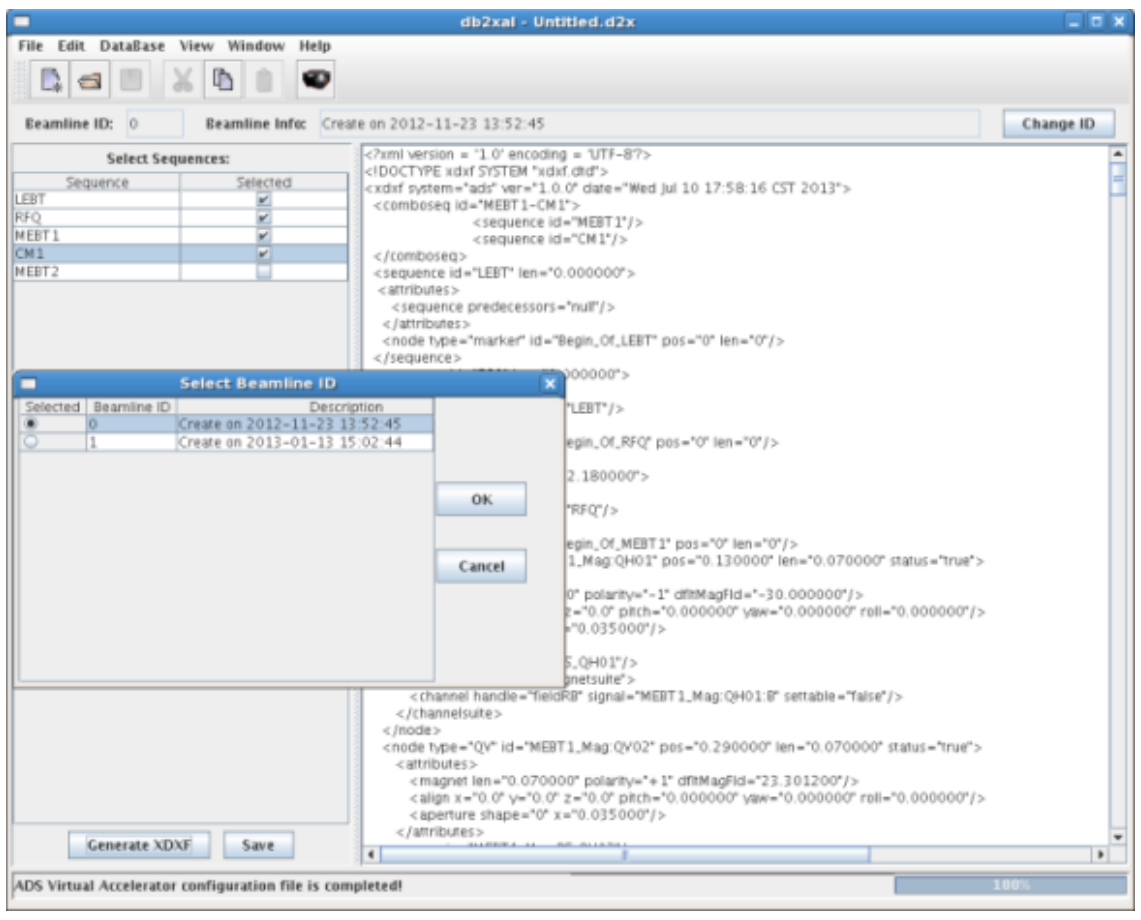

Fig.5: The main interface of db2xal

## 5. Virtual accelerator

With the Virtual Accelerator, it is possible to for an operator to judge whether setting parameters would be justified or not, examine the control system of the machine and practice the commissioning without a beam. The original virtual accelerator application don't support solenoid PVs and ADS new attribute—ETL, so proper modification is needed in order to better suit ADS online modeling purpose.

### 5.1 Modification

1) Adding support for solenoid PVs

• Modifying the registerNodeChannels method in VAServer.java to register PVs for the node type of solenoid

2) Adding new attribute—ETL

• Adding new attribute—ETL to the RfGapBucket class which define a set of RF gap attributes

• Adding definitions and functions in the RfGap class to handle the new attribute of ETL

• Modifying the definition of METHOD_LIVE_ETL and METHOD_DESIGN_ETL in the RfGapPropertyAccessor class

• Modifying the putSetPVs method, the configureReadbacks method and the SelectedSequenceChanged method in VADocument.java to suit the new attribute of ETL

### 5.2 Program test result

The main process of the ADS virtual accelerator will be in recurrent state after user start the simulation. The build-in EPICS portable channel access server broadcast all PVs which defined in the selected sequence or sequence combo within the LAN. Any computers in the same network segment can get and set the value of virtual PVs. The final result shown in Fig.7 validates that solenoid PVs and the new PV—cavETL can work normally. It also proves that the ADS configuration files generated by the application of db2xal are correct and the ADS virtual accelerator runs normally.

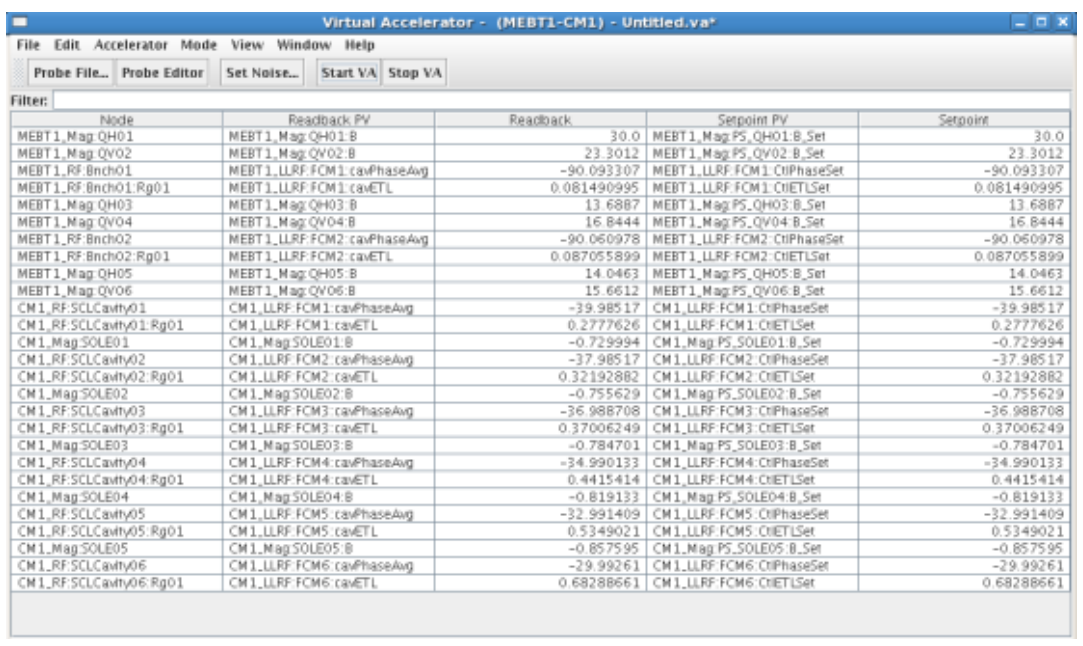

Fig.6: The main interface of virtual accelerator

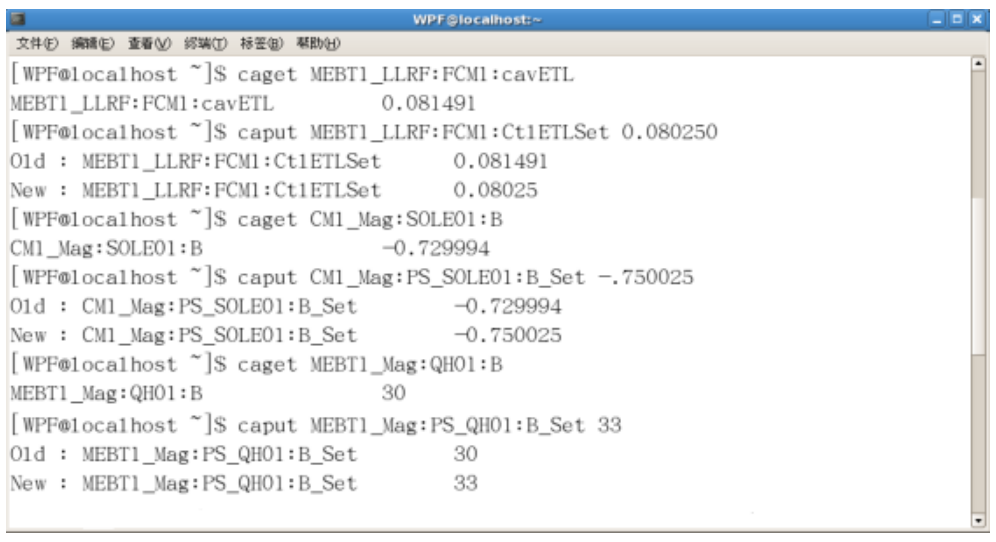

Fig.7: Test result of virtual accelerator